\documentclass[twocolumn,pre,aps,showpacs]{revtex4}
\usepackage{amsmath}
\usepackage{epsfig}

\begin{document}

\title{Topology hidden behind the breakdown of adiabaticity}
\author{Li-Bin Fu$^{1,2}$ and Shi-Gang Chen$^1$}
\affiliation{$^1$Institute of Applied Physics and Computational Mathematics, P.O. Box
8009 (28), 100088 Beijing, China,}
\affiliation{$^2$Max-Planck-Institute for the Physics of Complex systems, N\"{o}thnitzer
Str. 38, 01187 Dresden, Germany}

\begin{abstract}
For classical Hamiltonian systems, the adiabatic condition may fail at some
critical points. However, the breakdown of the adiabatic condition does not
always make the adiabatic evolution be destroyed. In this paper, we suggest
a supplemental condition of the adiabatic evolution for the fixed points of
classical Hamiltonian systems when the adiabatic condition breaks down at
the critical points. As an example, we investigate the adiabatic evolution
of the fixed points of a classical Hamiltonian system which has a number of
applications.
\end{abstract}

\maketitle

%\date{\today}

\section{Introduction}

Adiabaticity is an interesting concept in physics both for
theoretical studies and experimental practices
\cite{book,book1,book3,adiab,adiab1}. According to the adiabatic
theorem \cite{book}, if the parameters of the system vary with
time much more slowly than the intrinsic motion of the system, the
system will undergo the adiabatic evolution. For a classical
system, the adiabatic evolution means that the action of the
trajectory keeps invariant. For a quantum system, an initial
nondegenerate eigenstate remains to be an instantaneous eigenstate
when the Hamiltonian changes slowly compared to the level spacings
\cite{book}. Hence, the adiabatic evolution has been employed as
an important method of preparation and control of quantum states
\cite{adiaba,bergmann,raizen,comp}.

However, a problem may arise when the eigenstates become accident degenerate
at a critical point, i.e., when the level spacing tends to zero at a
critical point. For a classical system it corresponds to that the frequency
of the fixed point is zero at the critical point. The adiabatic condition is
not satisfied at the critical point because the typical time of the
intrinsic motion of the system becomes infinite. Can adiabatic evolution
still hold up when the adiabatic condition breaks down at the critical point?

Our motivation, derives from practical applications in current
pursuits of adiabatic control of Bose Einstein condensates (BECs)
\cite{bec}, which can often be accurately described by the
nonlinear Schr\"{o}dinger equation. Here the nonlinearity is from
a mean field treatment of the interactions between atoms.
Difficulties arise not only from the lack of unitarity in the
evolution of the states but also from the absence of the
superposition
principle. This was recently addressed for BECs in some specific cases \cite%
{band,kivshar}. But then, however, for such systems, only finite number of
levels are concerned. The nonlinear Schr\"{o}dinger equation of the system
with finite number of levels can be translated into a mathematically
equivalent classical Hamiltonian system. The evolution of an eigenstate just
corresponds to the evolution of a fixed point of the classical Hamiltonian
system. Then, the accident degeneracy of eigenstates is just translated into
accident collision of the fixed points. The latter one is quite well-known
subject and has been studied widely at least as a purely mathematical
problem \cite{book2}. Hence, our concern here is only focused on the
adiabatic evolution of the fixed points of classical Hamiltonian systems.

In this paper, we present a supplemental condition of the adiabatic
evolution for the fixed points of classical Hamiltonian systems when the
adiabatic condition breaks down at some critical points in the terms of
topology. As an example, we investigate the adiabatic evolution of the fixed
points of a classical Hamiltonian system which has a number of practical
interests. We show that the adiabatic condition will break down at
bifurcation points of the fixed points. But the adiabatic evolution is
destroyed only for the limit point. For the branch process, the adiabatic
evolution will hold, and the corrections to the adiabatic approximation tend
to zero with a power law of the sweeping rate.

\section{Supplemental adiabatic condition for the fixed points of classical
Hamiltonian systems}

For clarity and simplicity, we consider a one-freedom classical Hamiltonian $%
H(p,q;\lambda )$ with canonically conjugate coordinates $(p,q)\;$where $%
\lambda $ is a parameter of this system. The equations of motion are:

\begin{equation}
\overset{.}{q}=\frac{\partial H}{\partial p},\;\overset{.}{p}=-\frac{%
\partial H}{\partial q}.  \label{fpp}
\end{equation}
We can find two kinds of trajectories in the phase space for the system:
fixed points and closed orbits. The fixed points are the solutions of Eqs. (%
\ref{fpp}) when the right hands of them are zero. For a Hamiltonian system
there are only two kinds of the fixed points: elliptic points\ (stable fixed
points), hyperbolic points (unstable fixed points). The closed orbits are
around each of the elliptic points. We denote the fixed points by $%
z_{i}^{\ast }(p,q)\;(i=1,2,\cdots ,l)$ where $l$ is the total number of the
fixed points.

The action of a trajectory is defined as
\begin{equation}
I=\frac{1}{2\pi }\oint pdq,
\end{equation}
where the integral is along the closed orbit. Obviously, the action of a
fixed point is zero. The action is invariant when system undergos adiabatic
evolution.

According to the adiabatic theorem \cite{book}, the adiabatic condition can
be expressed as
\begin{equation}
\frac{2\pi }{\Omega }\frac{d\lambda }{dt}<<1,  \label{cond}
\end{equation}
where $\Omega $ is the frequency of the fixed point. If this condition
holds, the system will undergo adiabatic evolution, and keep the action not
varying. If $\Omega \neq 0,$ the condition can always be satisfied.

We can obtain the frequencies of the fixed points by linearized the
equations of motion. Let us define the Jacobian matrix as
\begin{equation}
J=\left(
\begin{array}{cc}
\frac{\partial ^{2}H}{\partial q^{2}} & \frac{\partial ^{2}H}{\partial
q\partial p} \\
\frac{\partial ^{2}H}{\partial q\partial p} & \frac{\partial ^{2}H}{\partial
p^{2}}%
\end{array}
\right) .  \label{jaco}
\end{equation}
It is well-known that when $\det (J)|_{z^{\ast }}>0,$ the fixed point is a
stable fixed point (elliptic point); when $\det (J)|_{z^{\ast }}<0,$ the
fixed point $z^{\ast }$ is a unstable fixed point (hyperbolic point).\ The
point with $\det (J)|_{z^{\ast }}=0$ is a degenerate point at which the
stability of the system is not determined.

For a stable fixed point $z^{\ast },$ the frequency of this fixed point is
\begin{equation}
\Omega _{0}=\sqrt{|\det (J)|_{z^{\ast }}}.  \label{fe}
\end{equation}
Obviously, $\Omega _{0}$ depends on the parameter $\lambda .$

Supposing at a critical point, namely $\lambda =\lambda _{c},$ we have $%
\Omega _{0}(\lambda _{c})=0.$ Therefore, the condition (\ref{cond}) will
break down at the point. We want to know what will happen when the adiabatic
condition fails (will the adiabatic evolution of the fixed point be
destroyed when the adiabatic condition does not hold ?).

In fact, if $|\det (J)|_{z^{\ast }}=0,$ the point $z^{\ast }$ is a
bifurcation point at which the fixed point will collide with the other fixed
points \cite{book2,fu}. Hence, the breakdown of adiabatic condition is
equivalent to collision of the fixed points (equivalent to accident
degeneracy of eigenstates of a corresponding quantum system). In the
collision process, fixed points may annihilate or merge into a stable fixed
point. The collision of the fixed points can be described clearly in the
terminology of topology \cite{fu}.

The equations of motion just define a tangent vector field $\mathbf{\phi }%
(p,q)=(\frac{\partial H}{\partial p},-\frac{\partial H}{\partial q})$ on the
phase space. Obviously, the fixed points $z_{i}^{\ast }(i=1,2,\cdots l)$ are
the zero points of the vector field, i.e., $\mathbf{\phi }(z^{\ast })=0$. We
know that the sum of the topological indices of the zero points of the
tangent vector field is the Euler number of the phase space which is a
topological invariant \cite{lisheng}. For a Hamiltonian system, the
topological index for a stable fixed point is $+1$ and for a unstable fixed
point is $-1.$

Indeed, if the fixed point is a regular point (not a degenerated point),
i.e., $\det (J)|_{z^{\ast }} \neq 0$, the topological index of the fixed
point can be determined by determinant of the Jacobian matrix defined by Eq.
(\ref{jaco}) \cite{lisheng,fu}. If $\det (J)|_{z^{\ast }}>0$, $z^{\ast }$ is
a stable fixed point and the topological index is $+1$; if $\det
(J)|_{z^{\ast }}<0,$ it is a unstable fixed point and the index is $-1.$

If $\det (J)|_{z^{\ast }}=0,$ i.e. if $z^{\ast }$ is a bifurcation point,
the topological index of this point seems to be not determined. As we have
shown before, the point is just the critical point of adiabatic evolution,
corresponding to collision of the fixed points.

However, because the sum of the topological indices is a topological
invariant, the topological index is conserved in a collision process of the
fixed points. Therefore, the topological index of the bifurcation point can
be determined by the sum of the indices of the fixed points involved in
collision. So, if the topological index of the bifurcation point is not
zero, it is still a fixed point after collision. But if the topological
index of the bifurcation point is zero, the bifurcation point will not be a
fixed point after collision.

Now, let us imagine what will happen when a fixed point is destroyed by a
collision process. Because there are only two kinds of trajectories for a
classical Hamiltonian system: fixed points and closed orbits around each of
the stable fixed point, so when a fixed point is destroyed, it will form a
closed orbit around the nearest stable fixed point. The action of the new
orbit must be proportional to the distance between the critical point and
the nearest stable fixed point. This sudden change of action (from zero to
finite) is so-called "adiabatic tunneling probability" which has been
studied in Refs. \cite{wuniu, add}. On the other hand, if the topological
index of the bifurcation point is $-1$, i.e., it is a unstable fixed point
after the collision, we can not expect the adiabatic evolution can keep on
after collision.

But if the topological index of the bifurcation point is $+1$, i.e., it is
still a stable fixed point after the collision, or in other word, the stable
fixed point survive after collision. For such case, the adiabatic evolution
will not be destroyed.

From above discussion, it is clear that when the adiabatic condition given
by Eq. (\ref{cond}) does not hold at a critical point with $\Omega _{0}=%
\sqrt{|\det (J)|_{z^{\ast }(\lambda _{c})}}=0,$ the system will still
undergo the adiabatic evolution if the topological index of the fixed point $%
z_{\ast }(\lambda _{c})$ is $+1$. On the contrary, if the topological index
of the point $z_{\ast }(\lambda _{c})$ is zero or $-1$ the adiabatic
evolution will be destroyed.

Hence, we get a supplemental condition of the adiabatic evolution of the
fixed points for a classical Hamiltonian system when the adiabatic condition
breaks down at a critical point. When the adiabatic condition is not
satisfied at a critical point, the topological property of the bifurcation
point plays an important role to judge whether the system will undergo
adiabatic evolution over this critical point: if the index of the
degenerated point $z_{\ast }(\lambda _{c})$ is $+1$ the adiabatic evolution
will hold. If the index of the point $z_{\ast }(\lambda _{c})$ is zero or $%
-1 $, the adiabatic evolution will not hold.

\section{A paradigmatic example and application}

As a paradigmatic example, we consider the following system

\begin{equation}
H(z,\theta ,\lambda ,\gamma )=-\sqrt{1-z^{2}}\cos \theta -\frac{\lambda }{2}%
z^{2}+\gamma z,  \label{ham}
\end{equation}
in which $(z,\theta )$ are canonically conjugate coordinates, and $\lambda $%
, $\gamma $ are two parameters. The equations of motion are : $\dot{\theta}=%
\frac{\partial H}{\partial z},\;\dot{z}=-\frac{\partial H}{\partial \theta }$%
, these yield
\begin{equation}
\dot{\theta}=\frac{z}{\sqrt{1-z^{2}}}\cos \theta -\lambda z+\gamma ,
\label{eq1}
\end{equation}
\begin{equation}
\dot{z}=-\sqrt{1-z^{2}}\sin \theta .  \label{eq2}
\end{equation}
This classical system can be obtained from a quantum nonlinear two-level
system, which may arise in a mean-field treatment of a many-body system
where the particles predominantly occupy two energy levels. For example,
this model arises in the study of the motion of a small polaron \cite{liufu6}%
, a Bose-Einstein condensate in a double-well potential \cite%
{liufu7,liufu8,liufu9} or in an optical lattice \cite{liufu10,liufu11}, or
for two coupled Bose-Einstein condensates \cite{lee,fupla}, or for a small
capacitance Josephon junction where the charging energy may be important.
This quantum nonlinear two-level model has also been used to investigate the
spin tunneling phenomena recently \cite{liufub}.

The fixed points of the classical Hamiltonian system are given by the
following equations
\begin{equation}
\theta ^{\ast }=0,\pi ,\;\;\;\gamma -\lambda z^{\ast }+\frac{z^{\ast }}{%
\sqrt{1-z^{\ast 2}}}\cos \theta ^{\ast }=0.  \label{fix}
\end{equation}
The number of the fixed points depends on the nonlinear parameter $\lambda .$
For weak nonlinearity, $\lambda \leq 1,$ there exist only two fixed points,
corresponding to the local extreme points of the classical Hamiltonian. They
are elliptic points located on lines $\theta ^{\ast }=$ $0$ and $\pi $
respectively, each being surrounded by closed orbits. For strong
nonlinearity, $\lambda >1,$ two more fixed points appear on the line $\theta
^{\ast }=$ $0$ in the windows $-\gamma _{c}<\gamma <\gamma _{c},$ one is
elliptic and the other one is hyperbolic as a saddle point of the classical
Hamiltonian, where $\gamma _{c}=(\lambda ^{2/3}-1)^{3/2}.$ In the following,
we only consider the cases in the region $-\gamma _{c}<\gamma <\gamma$.

We can obtain the frequencies of the fixed points by linearized the Eqs. (%
\ref{eq1}) and (\ref{eq2}). For the elliptic fixed points on line $\theta
^{\ast }=$ $0,$ the frequencies are equal, they are
\begin{equation}
\Omega _{0}=\sqrt{1-\lambda (1-z^{\ast 2})^{3/2}}.  \label{fre}
\end{equation}
Obviously, if $z^{\ast }=z_{c}^{\ast }=\sqrt{1-\left( \frac{1}{\lambda }%
\right) ^{2/3}},$ the frequencies will be zero, i.e., $\Omega
_{0}(z_{c}^{\ast })=0$. From Eq. (\ref{fix}), we can obtain, when $\lambda
=\lambda _{c}=$ $(\gamma ^{2/3}+1)^{3/2},$ one of the elliptic fixed point
will be $(z^{\ast }=z_{c}^{\ast },\theta ^{\ast }=0).$ At this point the
adiabatic condition will break down.

Hence, if we start from this elliptic fixed point on line $\theta ^{\ast }=$
$0$ at $\lambda =\lambda _{0}>1$, and $\lambda $ changes with time as $%
\lambda =\lambda _{0}-\alpha t$ (keeping $\gamma $ invariant in the window $%
-\gamma _{c}(\lambda _{0})<\gamma <\gamma _{c}(\lambda _{0})$)$,$ the
adiabatic condition (\ref{cond}) will break down at the point $z_{c}^{\ast }$
when $\lambda $ reaches $\lambda _{c}$ because $\Omega _{0}=0$. We want to
know what will happen when the adiabatic condition is not satisfied (will
the adiabatic evolution be destroyed when the adiabatic condition does not
hold ?). There are two different cases for discussing: $\gamma \neq 0$ and $%
\gamma =0.$

%%%%%%%%%%%%%%%%%%%%%%%%%%%%%%%%%%%%%%%%%%%%%%%%%%%%%%%%%%%%%%%%%%%%%%%%%%%%%%%%%%%%%%%%%
\begin{figure}[!htb]
\begin{center}
\rotatebox{-90}{\resizebox *{7cm}{8.0cm} {\includegraphics
{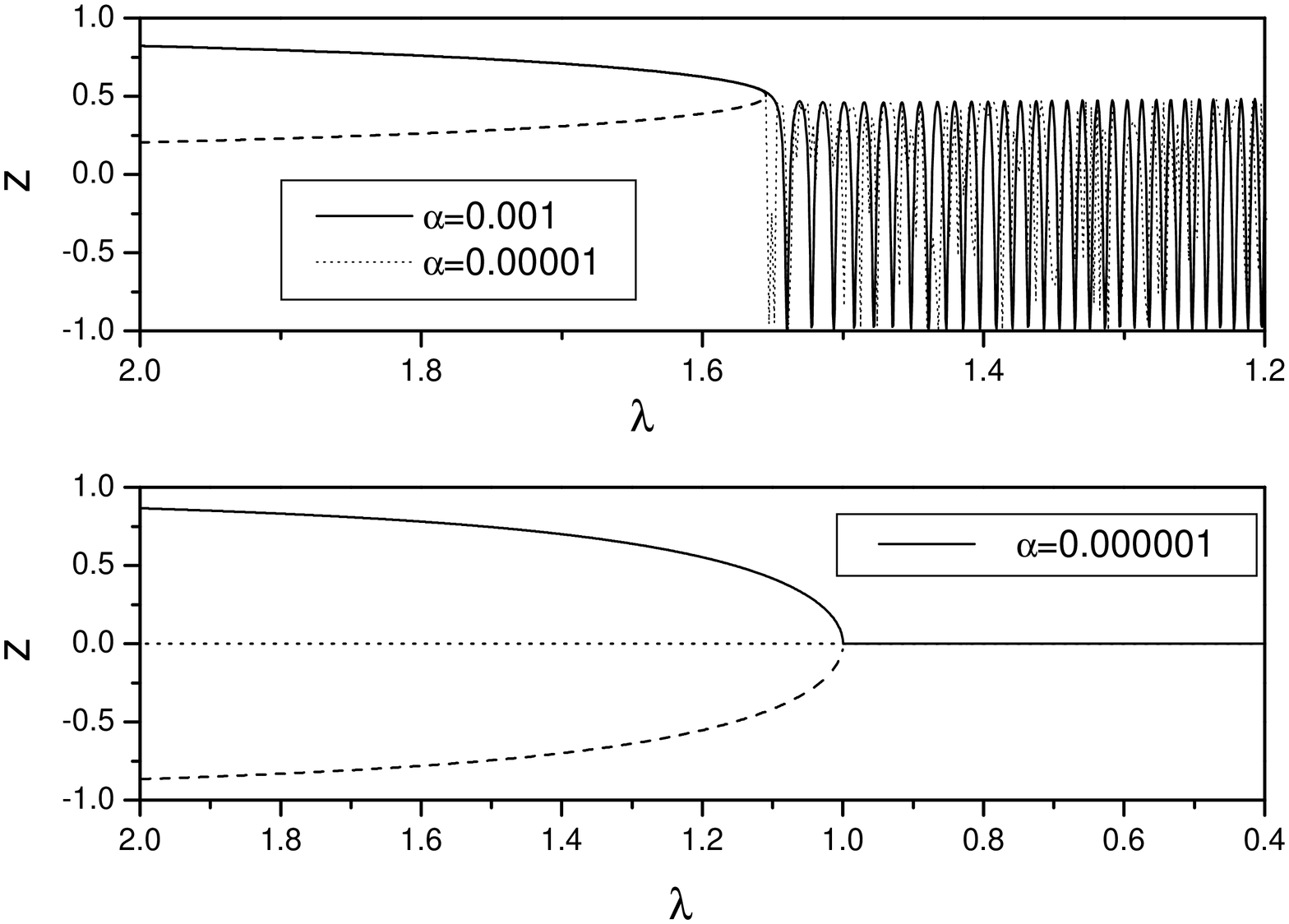}}}
\end{center}
\caption{Time evolution of $z(t)$ initially on the elliptic fixed point $%
z^{\ast }(\protect\lambda _{0})$. (a) for $\protect\gamma =0.2,$ $\protect%
\lambda _{0}=2.0.$ The solid line denotes the time evolution of $z(t)$ for $%
\protect\alpha =0.001$ and dotted line for $\protect\alpha =0.00001$. The
dashed line refers to the trajectory of the hyperbolic point. (b) for $%
\protect\gamma =0,$ $\protect\lambda _{0}=2.0.$ The solid line denotes the
time evolution of $z(t)$ for $\protect\alpha =0.000001.$ The dotted line
refers to the trajectory of the hyperbolic point. The dashed line refers to
the trajectory of another elliptic point.}
\label{fig1}
\end{figure}
%%%%%%%%%%%%%%%%%%%%%%%%%%%%%%%%%%%%%%%%%%%%%%%%%%%%%%%%%%%%%%%%%%%%%%%%%%%%%%%%%%%%%%%%%%

Case 1 $(\gamma \neq 0)$: For the convenience, we choose $\gamma =0.2$ and $%
\lambda _{0}=2$. We start at the elliptic fixed point $(z^{\ast }=z^{\ast
}(\lambda _{0}),\theta ^{\ast }=0)$ and $\lambda $ varies with very small $%
\alpha .$ At the beginning, the system follows the $z^{\ast }[\lambda (t)]$
adiabatically. But when $\lambda $ reaches $\lambda _{c}=$ $(\gamma
^{2/3}+1)^{3/2},$ the adiabatic evolution is destroyed with a jump of action
(the action changes to a finite value from zero suddenly) at the point $%
z_{c}^{\ast }\;(\approx 0.5048)$. Fig. 1(a) shows this process. Obviously,
the breakdown of adiabatical condition leads to the destroy of the adiabatic
evolution.

Case 2 $(\gamma =0)$: From Eq. (\ref{ham}) and (\ref{fix}), we can have two
elliptic fixed points on line $\theta ^{\ast }=$ $0$ for $\lambda >1$,
\begin{equation}
z_{\pm }^{\ast }=\pm \sqrt{1-1/\lambda ^{2}},\;\theta ^{\ast }=0,\;\Omega
_{0}=\sqrt{\lambda (\lambda ^{2}-1)},  \label{l01}
\end{equation}%
and for $\lambda \leq 1,$ there is only one fixed point,
\begin{equation}
z^{\ast }=0,\;\theta ^{\ast }=0,\;\Omega _{0}=\sqrt{1-\lambda }.  \label{l02}
\end{equation}%
Obviously, for $z_{c}^{\ast }=0$ and $\lambda _{c}=1.$, $\Omega _{0}=0,$ so
adiabatic condition can not be satisfied. We integrate the classical
equations of the Hamiltonian system (\ref{ham}), with the initial condition $%
\lambda _{0}=2,$ $z_{0}=z_{+}^{\ast }(\lambda _{0})$, and $\theta ^{\ast
}(0)=0.$ Fig. 1(b) shows the time evolution of this fixed point for a very
small sweeping rate $\alpha $. The final state is a very small oscillation
around the fixed point $(z^{\ast }=0,\theta ^{\ast }=0)$. In Fig. 2, we plot
the dependency of the small oscillation amplitude $\delta $ on the sweeping
rate $\alpha $. From this figure, it is clear to see that the amplitude of
the small oscillation will tend to zero with the sweeping rate decreasing as
a power law: $\delta =0.7\ast \alpha ^{1/2}$. Therefore, for this case, the
system will evolve adiabatically and keep the action not changing for all
the time if the sweeping rate is small enough, even when $\lambda $ crosses
the critical point $\lambda _{c}=1,$ at which $\Omega _{0}(z_{c}^{\ast })=0$%
, i.e., though the adiabatic condition is not satisfied when $\lambda $
crosses the point $\lambda _{c}=1$, the system is still undergoing adiabatic
evolution.

%%%%%%%%%%%%%%%%%%%%%%%%%%%%%%%%%%%%%%%%%%%%%%%%%%%%%%%%%%%%%%%%%%%%%%%%%%%%%%%%%%%%%%%%%
\begin{figure}[!htb]
\begin{center}
\rotatebox{-90}{\resizebox *{6cm}{7.5cm} {\includegraphics
{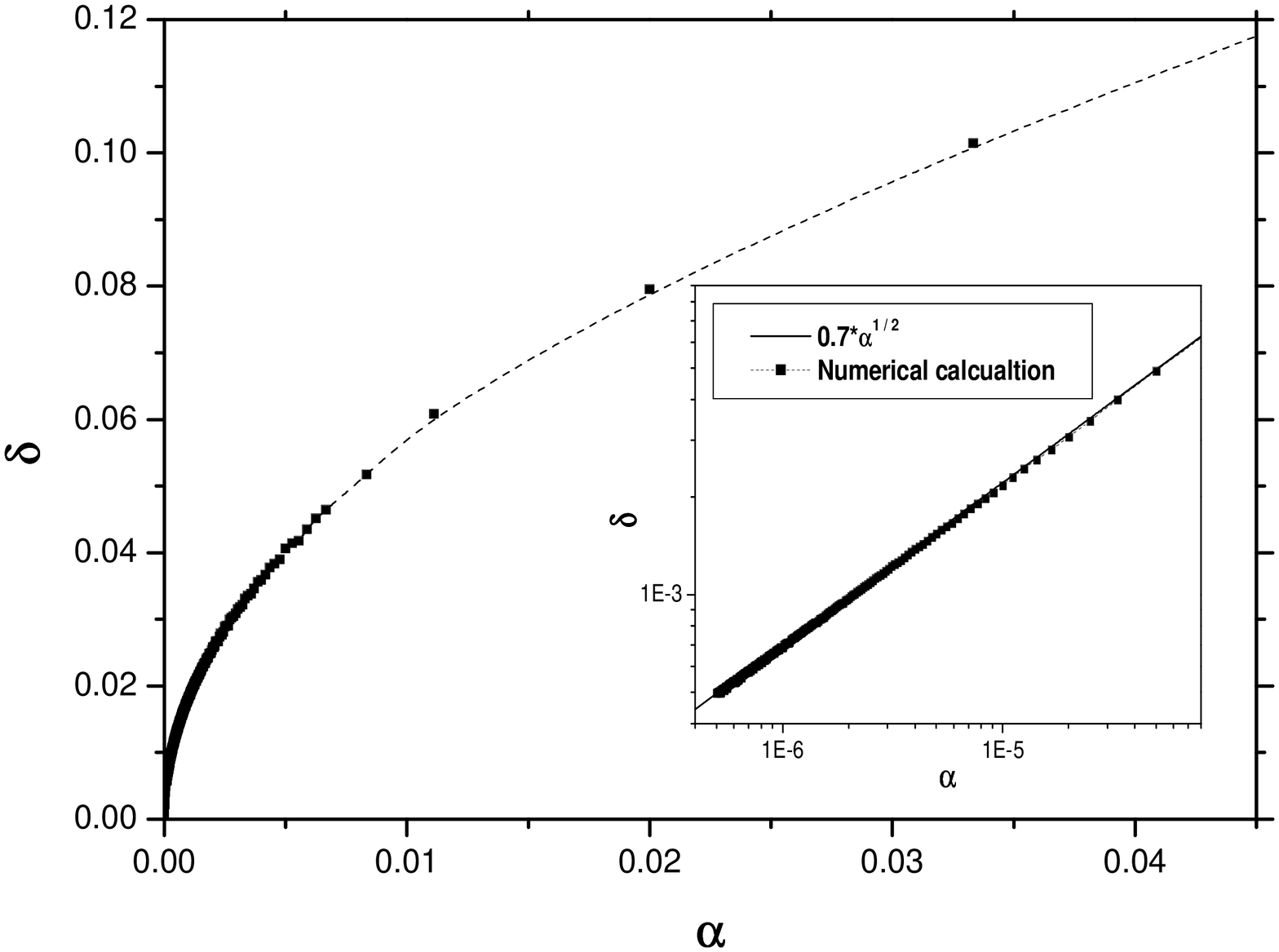}}}
\end{center}
\caption{The dependency of the oscillation amplitude of the final
state on the sweeping rate $\protect\alpha$ for the Case 2. The
initial state and parameters are the same as the Fig 1 (b) except
the sweeping rate. The solid quadrangles are
our numerical calculations. The solid line of insert figure is the function $%
0.7*\protect\alpha^{1/2}$.}
\label{fig2}
\end{figure}
%%%%%%%%%%%%%%%%%%%%%%%%%%%%%%%%%%%%%%%%%%%%%%%%%%%%%%%%%%%%%%%%%%%%%%%%%%%%%%%%%%%%%%%%%%

In fact, if we make the series expansion of the Hamiltonian (\ref{ham})
around the critical point, the system can be approximated to a double well
system \cite{liufu8}. Therefore, the phenomenon of Case 2 can be illustrated
by the standard double well model. Considering a particle in a double well,
the system is described by the Hamiltonian $H=1/2p^{2}-1/2\mu x^{2}+1/4x^{4}.
$ For $\mu >0,$ it has two stable fixed points $(x,p)=(\sqrt{\mu },0)$ and $%
(x,p)=(-\sqrt{\mu },0),$ and an unstable fixed point $(x,p)=(0,0);$ for $\mu
<0$ it has a single stable fixed point $(x,p)=(0,0)$. At the critical point $%
\mu =0$, three fixed points merge into a stable fixed point. As the
parameter $\mu $ varies from $+1$ to $-1$ the system goes from a double well
to a single well. The stable fixed points are just the bottom of the wells,
and the unstable point is just the saddle point of double well. If the
particle is at the fixed point $(\sqrt{\mu },0)$ at the beginning, i.e., the
particle stays at the bottom of one well. Then, let $\mu $ vary very slowly.
At the critical point $\mu =0,$ the two wells merge into a single well. At
this time, the bifurcation point is $(x,p)=(0,0),$ which is the bottom of
the single well. So if $\mu $ varies very slowly, one can imagine that the
particle will stay at the bottom of the well all the time, even when the
system goes from a double well to a single one (at this time the adiabatic
condition does not hold but the bifurcation point is still a stable fixed
point, because the bifurcation point still corresponds to the bottom of the
well).

As we have discussed in Sect. II, the breakdown of the adiabatic condition $%
(\Omega _{0}=0)$ corresponds to the trajectory bifurcation, i.e., the points
$z_{c}^{\ast }$ is just a bifurcation point of the fixed points. The
properties of the fixed points are determined by the following Jacobian
\begin{equation}
D=\left. \det \left(
\begin{array}{cc}
\frac{\partial \phi ^{1}}{\partial z} & \frac{\partial \phi ^{2}}{\partial z}
\\
\frac{\partial \phi ^{1}}{\partial \theta } & \frac{\partial \phi ^{2}}{%
\partial \theta }%
\end{array}
\right) \right| _{(z^{\ast },0)}.  \label{deter}
\end{equation}
Where $\mathbf{\phi }(z,\theta)=(\frac{\partial H}{\partial z},-\frac{%
\partial H}{\partial \theta})$. If the Jacobian $D\neq 0,$ the zero point
(fixed point) is a regular point. But when $D=0$, the zero point is a
bifurcation point.

There are two kinds of bifurcation points: limit points and branch points.
The limit point satisfies that $D=0$ but $D^{1}=\left. \det \left(
\begin{array}{cc}
\frac{\partial \phi ^{1}}{\partial \lambda } & \frac{\partial \phi ^{2}}{
\partial \lambda } \\
\frac{\partial \phi ^{1}}{\partial \theta } & \frac{\partial \phi ^{2}}{
\partial \theta }%
\end{array}
\right) \right| _{(z_{c}^{\ast },0)}$ $\neq 0,$ which corresponds to
generation and annihilation of the fixed points.

If $D|_{(z_{c}^{\ast },0)}=D^{1}|_{(z_{c}^{\ast },0)}=0$, the point $%
(z_{c}^{\ast },0)$ is a branch point. The branch point corresponds to branch
process of the fixed points. The directions of all branch curves are
determined by the equations \cite{fu}
\begin{equation}
A\frac{d^{2}z}{d\zeta ^{2}}+2B\frac{dz}{d\zeta }+C=0,  \label{d1}
\end{equation}
or
\begin{equation}
C\frac{d^{2}\zeta }{dz^{2}}+2B\frac{d\zeta }{dz}+A=0,  \label{d2}
\end{equation}
where $A,$ $B,$ and $C$ are three constants. $\zeta $ corresponds to $%
\lambda $ or $\ \gamma $ respectively.\ Different solutions of the above
equations correspond to different branch processes.

For the zero point $(z^{\ast },0)$, i.e., the fixed point on line $\theta
^{\ast }=0,$ we can obtain $D=\left. -\sqrt{1-z^{\ast 2}}\Omega
_{0}^{2}\right| _{(z^{\ast },0)}.$ \ Obviously, when $\Omega _{0}=0$, $D=0,$
the critical point $(z_{c}^{\ast },0)$ is a bifurcation point, at which the
adiabatic condition fails.

For the case 1: We can find at the point $z_{c}^{\ast },$ the Jacobian $%
D=\left. -\sqrt{1-z^{2}}\Omega _{0}^{2}\right| _{(z_{c}^{\ast },0)}=0,$ but $%
D^{1}=\left. -s\sqrt{1-z^{2}}\right| _{(z_{c}^{\ast },0)}\neq 0.$ This point
is a limit point which corresponds to annihilating of zero points. At this
point, the elliptic point annihilates simultaneously with a hyperbolic
point. In Fig. 1(a), the dashed lines is the trajectory of the hyperbolic
point. Apparently, the elliptic point evolves adiabatically until it
annihilates with the hyperbolic point at $z_{c}^{\ast }.$ After this
annihilation the elliptic point turns to an ordinary closed orbit with a
nonzero action, so the adiabatic evolution is destroyed. The annihilation
process of the fixed points of the system (\ref{ham}) has also been
discussed in Ref. \cite{fu2} in detail.

For the case 2: At the point $z_{c}^{\ast }=0,$ the Jacobian determinant $%
D=0,$ and $D^{1}=\left. -z\sqrt{1-z^{2}}\right| _{(z_{c}^{\ast },0)}=0$.
This is a branch process of the fixed points. We can prove that for this
case $A=C=0$, so the solutions of equations (\ref{d1}) and (\ref{d2}) give
two directions: $\frac{dz}{d\lambda }=0,$ and $\frac{d\lambda }{dz}=0.$ The
branch process corresponds to merging process. At this branch point, three
fixed points, two elliptic points and one hyperbolic point, merge together.
One can see this point in Fig. 1(b), in which the dotted line is the
trajectory of the hyperbolic point, and the dashed line corresponds to the
trajectory of another elliptic point. Since the total topological index is
invariant, the three fixed points merge to one point with index $+1,$ i.e.,
merge to an elliptic point. The elliptic point evolves adiabatically until
it reaches the critical point $z_{c}^{\ast },$ at which three fixed points
merges to one elliptic point. Therefore, after the branch process, the
elliptic point turns to a new elliptic point, the action keeps zero and the
adiabatic evolution still holds.

From above discussion, we see that the adiabaticity breaks down at
bifurcation points of the fixed points, but only for the limit point the
adiabatic evolution is destroyed (case 1), while for this case the two fixed
points annihilate. For case 2, three fixed points merge to one, because the
critical point $(z_{c}^{\ast },0)$ is still a stable fixed point, the
adiabatic evolution keeps with action zero.

The phenomena discussed above can occur for the adiabatic change of $\gamma$
with $\lambda$ fixed. On the other hand, the Hamiltonian (\ref{ham}) is
invariant under the transformations $\lambda \rightarrow -\lambda ,$ $\theta
\rightarrow \theta +\pi ,$ and $t\rightarrow -t.$ Hence, the phenomena can
also be found under such transformations.

\section{Conclusion}

In summary, at some critical points, the adiabatic condition fails, but the
adiabatic evolution may not always be broken. We find that the topological
property of the critical point plays an important role for adiabatic
evolution of the fixed points when the adiabatic condition does not hold. If
the topological index of the critical point is $+1$ the adiabatic evolution
of fixed point will not be destroyed. On the contrary, if the index of the
critical point is zero or $-1$, the adiabatic evolution will be destroyed.
As a paradigmatic example, we investigated the adiabatic evolution of a
classical Hamiltonian system which has a number of practical interests. For
this system, the adiabaticity breaks down at bifurcation points of the fixed
points. But only for the limit point the adiabatic evolution is destroyed.
For the branch process, the adiabatic evolution will hold, and the
corrections to the adiabatic approximation tend to zero with a power law of
the sweeping rate.

In general, the corrections to the adiabatic approximation are exponentially
small in the adiabaticity parameter, both for quantum system and classical
system \cite{book,book1,book3}. It is particularly interesting that the
corrections of the adiabatic approximation may be a power law (e.g., for the
case 2). The power law corrections to daidabatic approximation have also
been found in the nonlinear Landau-Zener tunneling \cite{wuniu}. In Ref. %
\cite{wuniu}, the authors found that when the nonlinear parameter is smaller
than a critical value, the adiabatic corrections are exponentially small in
the adiabatic parameter, but when the nonlinear parameter equals to the
critical value, the adiabatic corrections are a power law of the adiabatic
parameter. Furthermore, if the nonlinear parameter is larger than the
critical value, the so-called non-zero adiabatic tunneling will occur \cite%
{wuniu,add}. Indeed, the cases, for which the corrects to the adiabatic
approximation are not exponential law with the adiabatic parameter,
correspond to the collision of fixed points.

\section*{Acknowledgments}

This work was supported by the 973 Project of China and National
Nature Science Foundation of China (10474008,10445005). LB Fu is
indebted to Dr. Chaohong Lee and Alexey Ponomarev for reading this
paper, and acknowledges funding by the Alexander von Humboldt
Stiftung.

\end{document}